\documentclass[11pt,a4paper]{article}

\usepackage{a4wide}

\usepackage[margin=2cm]{geometry}
\usepackage[margin=1cm,font=small]{caption}

\usepackage{stix2}

\usepackage{amsmath,amssymb,amsthm}
\usepackage{bm}
\usepackage[hidelinks]{hyperref}
\usepackage[nameinlink]{cleveref}
\usepackage{enumitem}
\usepackage{pgf}
\usepackage{authblk}
\usepackage{microtype}

\title{Quantum electrodynamical formulation of photochemical acid generation and its implications on optical lithography}

\author{Seungjin Lee\thanks{\href{mailto:seungjin.lee@etri.re.kr}{seungjin.lee@etri.re.kr}}}
\affil[]{Electronics and Telecommunications Research Institute,\\ Gajeong-ro 218, Daejeon, Republic of Korea, 34129}

\begin{document}
\maketitle

\abstract{The photochemical acid generation is refined from the first principles of quantum electrodynamics. First, we briefly review the formulation of the quantum theory of light based on the quantum electrodynamics framework to establish the probability of acid generation at a given spacetime point. The quantum-mechanical acid generation is then combined with the deprotection mechanism to obtain a probabilistic description of the deprotection density directly related to feature formation in a photoresist. A statistical analysis of the random deprotection density is presented to reveal the leading characteristics of stochastic feature formation.}

\section{Introduction}
\label{sec:intro}

The physics of optical lithography has steadily entered the quantum regime owing to the constant demand to reduce the physical length scale of the patterning process. In particular, the classical-to-quantum transition has been accelerated by introducing an extreme ultraviolet (EUV) exposure system with a typical wavelength of approximately $10^{-8}$m. 
    
Among other things, the quantum effect in EUV lithography has been most faithfully represented by shot noise arising from the quantum fluctuation of the incident light intensity. Such quantum fluctuation has been statistically averaged out for an exposure system having a high photon density of states (i.e., dose per energy of each photon). However, for EUV exposure, no drastic increase in the exposure dose compared to the increment in individual photon energy gives rise to severe quantum fluctuations in light intensity.
    
It has been widely accepted that the fluctuations can be depicted by the randomness of the number of photons following a Poisson distribution under a given dose \cite{mack2010}. Nevertheless, the fully quantum-mechanical exposition of shot noise in the context of optical lithography has barely been studied.
    
This paper provides a firm physical ground for quantum-mechanical shot noise in optical lithography by employing quantum electrodynamics (QED) \cite{schwinger1948}. More precisely, we reformulate the photochemical process for acid generation in the language of QED by identifying the photochemical process with the photoelectric detection of light.
    
The QED formulation of acid generation merely demonstrates the well-established proportionality between classical light intensity and photoelectric detection probability \cite{mandel1964}. However, the exceptional merit of introducing QED to optical lithography is the capability to integrate the physical descriptions of primary and secondary photochemical processes, such as flares and secondary electrons, into a single framework, although such secondary effects are not discussed in this work.
    
With probabilistic acid generation, we proceed to the stochastic deprotection mechanism of the photoresist by defining deprotection density as a random field on the spacetime points of localized acid generation. The random deprotection density then allows us to compute the probability of pixel formation at each spatial location in the resist, which leads to the probability of a feature forming in a given region. 
    
The probabilistic feature formation in our work is closely related to the former works in \cite{gallatin2005, latypov2020}. However, our work elaborates on discussions in those works with detailed quantum-mechanical acid generation and rather refined statistical analyses.
    
The remainder of this paper is organized as follows. In the following section, we provide a QED formulation for photochemical acid generation, including a brief review of the construction of the QED and a revision of the photoelectric detection of light. In Section \ref{sec:dep}, we present a statistical analysis of the randomness of feature formation in a resist arising from the probabilistic nature of acid generation. Section \ref{sec:con} is devoted to the conclusions and outlook of the current work.

\section{Quantum-mechanical formulation of acid generation}
\label{sec:qed}
	
In optical lithography, a feature on a photoresist is formed by developing deprotected polymers in the photoresist using a resist developer. Deprotected polymers are generated through a deprotection mechanism carried out by the chemical reaction between protected polymers and acids, in which the latter are produced by the photochemical process of photoacid generators (PAGs). 
	
At the molecular level, acid generation occurs when the confined electrons of PAGs are released through interactions with the exposure light. Accordingly, the rate of acid generation can be identified with that of photoelectric detection of light, which is a well-known application of the quantum theory of light. The following subsections provide a quantum-mechanical description of acid generation by reformulating the photoelectric detection of light in QED.

\subsection{Quantum theory of light: a brief review}

The quantum theory of light and its interaction with charged particles has been most successfully formulated in the quantum field theory (QFT) framework, ensuring the manifest spacetime Lorentz symmetry and the unitarity of quantum mechanics. In this subsection, we briefly review the formulation of the QFT for light and charged particles, called quantum electrodynamics, following the notations and conventions of \cite{weinberg1995}\footnote{In particular, Planck's constant $\hbar$ and the speed of light $c$ are taken to be unity. We also use the Einstein summation convention for repeated indices.}.
    
In the QFT framework, the elementary quantum state relevant to light is an excited quantum state over the ground state $\vert 0 \rangle$, called the photon. Photons can be created and annihilated by the set of creation and annihilation operators $\left\{a\left(p, \sigma\right), a^{\dagger}\left(p, \sigma\right)\right\}$, where $p = \left(p^{0}, p^{1}, p^{2}, p^{3}\right)$ and $\sigma = \pm 1$ denote the four-vector momentum and intrinsic spin of the photon, respectively. 

The four-vector momentum $p$ satisfies the massless condition
\begin{align}
0 = \left(p^{1}\right)^{2} + \left(p^{2}\right)^{2} + \left(p^{3}\right)^{2} - \left(p^{0}\right)^{2}
\label{eq:masslessCondition}
\end{align}
together with $p^{0} \ge 0$, where $p^{0}$ is the photon energy. The creation-annihilation pair satisfies the following commutation relation:
\begin{align}
\left[ a\left(\bm{p},\sigma \right), a^{\dagger}\left(\bm{p}',\sigma'\right)\right] = \delta_{\sigma\sigma'} \delta\left(\bm{p} - \bm{p}'\right)
\end{align}
for $\bm{p} = \left(p^{1}, p^{2}, p^{3}\right)$ and $\bm{p}' = \left(p'{}^{1}, p'{}^{2}, p'{}^{3}\right)$.

The creation-annihilation pair forms the quantum field $A^{\mu}\left(x\right)$ ($\mu = 0,1,2,3$) via
\begin{align}
A^{\mu}\left(x\right) = A^{\mu}_{\left(+\right)}\left(x\right) + A^{\mu}_{\left(-\right)}\left(x\right)
\end{align}
where
\begin{subequations}
    \label{eq:Adecomposed}
    \begin{align}
        A^{\mu}_{\left(+\right)}\left(x\right)=& \left( 2 \pi \right)^{-3/2} \sum_{\sigma = \pm 1} \int \frac{d^{3}p}{\sqrt{2 p^{0}}} e^{\mu}\left(\bm{p},\sigma \right) a\left( \bm{p},\sigma \right) e^{ipx} \\
        A^{\mu}_{\left(-\right)}\left(x\right) =& \left( 2 \pi \right)^{-3/2} \sum_{\sigma = \pm 1} \int \frac{d^{3}p}{\sqrt{2 p^{0}}} e^{\mu*}\left( \bm{p},\sigma \right) a^{\dagger}\left( \bm{p},\sigma \right) e^{-ipx}
    \end{align}
\end{subequations}
with the polarization vector $e^{\mu}$ and its complex conjugate $e^{\mu *}$.

The massless condition in \eqref{eq:masslessCondition} is translated into the equations of motion for $A^{\mu}$ as
\begin{align}
\square A^{\mu} = 0
\end{align}
which is a homogeneous wave equation. Also, the Lorentz symmetry implies the dynamics of $A^{\mu}$ to be invariant under the gauge transformation given by
\begin{align}\label{eq:gaugeTr}
A^{\mu}\left(x\right) \mapsto A^{\mu} \left(x\right) + \partial^{\mu} \Omega\left(x\right), \quad \partial^{\mu} = \frac{\partial}{\partial x_{\mu}}.
\end{align}
    
To construct the dynamics of $A^{\mu}$ consistent with the gauge symmetry, we introduce the field strength 
\begin{align}
F_{\mu\nu} = \partial_{\mu} A_{\nu} - \partial_{\nu} A_{\mu}
\end{align}
showing the linear relation between the field strength $F$ and $A$\footnote{The linearity between a given quantum field and its field strength reflects the abelian nature of the gauge transformation in \eqref{eq:gaugeTr}. In addition, linearity ensures no self-interaction between photons; therefore, photons can only interact with each other through charged particles.}. The most general form of the action is given by
\begin{align}\label{eq:actionOfA}
S = \int d^{4} x \mathcal{L}, \quad \mathcal{L} = - \frac{1}{4} F_{\mu\nu} F^{\mu\nu} + J_{\mu} A^{\mu} + \mathcal{L}_{\text{matter}}
\end{align}
where $J_{\mu} = J_{\mu} \left(\psi\right)$ is the current written in some charged matter fields $\psi$ and $\mathcal{L}_{\text{matter}}$ is the Lagrangian for the charged particles represented by $\psi$.
    
The dynamical system of $A_{\mu}$ represented by \eqref{eq:actionOfA} is a constrained system owing to the gauge symmetry in \eqref{eq:gaugeTr}. Consequently, only $A_{m}$ ($m = 1,2,3$) are dynamical, and we have to choose an appropriate gauge to exhaust gauge freedom. Here, we choose the Coulomb gauge
\begin{align}
\partial_{m} A^{m} = 0, \quad m = 1, 2, 3.
\end{align}
    
Altogether, the Hamiltonian of the system is obtained as follows:
\begin{align}
H = \int d^{3}x \left( \frac{1}{2} E^{m} E_{m} + \frac{1}{2} B^{m}B_{m} + J^{m}A_{m} \right) + H_{\text{matter} \oplus \text{Coul}}
\label{eq:hamOfA}
\end{align}
where
\begin{align}
F^{0m} = E^{m}, \quad F^{mn} = \epsilon^{mnp} B_{p}, \quad m,n,p = 1,2,3
\end{align}
and
\begin{align}
H_{\text{matter} \oplus \text{Coul}} = H_{\text{matter}} - \frac{1}{2} \int d^{3}x d^{3}y \frac{J^{0}\left(x\right) J^{0}\left(y\right)}{4\pi \vert \bm{x} - \bm{y} \vert}.
\end{align}
Having the Hamiltonian $H$ in \eqref{eq:hamOfA} for light and charged matter, one can compute the evolution of a quantum state given at the initial time. A detailed discussion of the quantum evolution of a physical system can be found in the standard literature, such as \cite{messiah2014}.

\subsection{Photoelectric detection probability}
    
As indicated at the beginning of this section, acid generation can be identified by the photoelectric detection of light. In this process, the exposure system can be represented by the Hilbert space $\mathcal{H}$, which is the product space $\mathcal{H}_{\gamma} \otimes \mathcal{H}_{\psi}$ of the space of photons $\mathcal{H}_{\gamma}$ and charged particles $\mathcal{H}_{\psi}$. At the initial time $t_{0}$ two systems are completely disentangled \cite{mandel1995}, so that the initial state $\vert \Psi\left(t_{0}\right) \rangle \in \mathcal{H}$ is given by
\begin{align}
\vert \Psi\left(t_{0}\right) \rangle = \left\vert\chi \right\rangle \otimes \left\vert \psi\right\rangle, 
\quad \left\vert\chi \right\rangle \in \mathcal{H}_{\gamma} \text{ and } \left\vert \psi \right\rangle \in \mathcal{H}_{\psi}.
\end{align}
    
For the detection of light, we are only interested in the occurrence of an interaction between light and charged matter at a given time $t>t_{0}$. Accordingly, we do not need to specify the states in $\mathcal{H}_{\psi}$ at $t_{0}$ and $t$ if they satisfy the following conditions:
\begin{align}
\left\vert\psi\right\rangle \perp \left\vert\bar{\psi}\right\rangle\quad \Leftrightarrow\quad \left\langle \bar{\psi}\right.\left\vert \psi \right\rangle = 0,\quad \left\vert\psi\right\rangle, \left\vert\bar{\psi}\right\rangle \in \mathcal{H}_{\psi}.
\end{align}
where $\left\vert\bar{\psi}\right\rangle$ denotes the state of charged matter at time $t$. Also, we trace $\mathcal{H}_{\gamma}$ out at $t$ since the final state of light is not a concern. 
    
Altogether, we have the transition amplitude $T\left(t, t_{0}\right)$ equivalent to the probability of a photochemical interaction as follows:
\begin{align}
T\left(t,t_{0}\right) = \sum_{\substack{\psi, \bar{\psi} \in \mathcal{H}_{\psi}\\\psi\perp \bar{\psi}}}\text{Tr}\, \left( \left\vert \bar{\psi} \right\rangle \left\langle \bar{\psi} \right\vert \rho_{\psi} \left(t\right) \right)
\end{align}
where $\rho_{\psi}\left(t\right)$ denotes the density matrix obtained by tracing out $\mathcal{H}_{\gamma}$ as
\begin{align}
\rho_{\psi}\left(t\right) = \underset{\mathcal{H}_{\gamma}}{\text{Tr}}\, \rho\left(t\right), 
\quad \rho\left( t \right) = \left\vert \Psi\left( t \right) \right\rangle \left\langle \Psi\left( t \right) \right\vert.
\end{align}
In perturbative expansion, the first nonvanishing contribution to $T\left(t, t_{0}\right)$ is given by \cite{mandel1995}
\begin{align}
T\left(t, t_{0}\right) \simeq& \sum_{\substack{\psi, \bar{\psi} \in \mathcal{H}_{\psi}\\\psi\perp \bar{\psi}}} \int_{t_{0}}^{t} d^{4}x_{2} \int_{t_{0}}^{t_{2}} d^{4}x_{1}  T^{(2)}(\psi,\bar{\psi},J,A;x_{1},x_{2}) + c.c.
\label{eq:transMaster}
\end{align}
where $c.c.$ denotes the complex conjugate of the previous term and
\begin{align}
    T^{(2)}(\psi,\bar{\psi},J,A;x_{1},x_{2}) = 
    &\left\langle \bar{\psi} \right \vert J_{m} \left(x_{1}\right) \left\vert \psi \right\rangle \left\langle \psi \right\vert J_{n} \left(x_{2}\right) \left\vert \bar{\psi} \right\rangle
    \left\langle \chi \right\vert A^{m}\left(x_{1}\right) A^{n}\left(x_{2}\right) \left\vert \chi \right\rangle.
\end{align}

To simplify the transition amplitude in \eqref{eq:transMaster}, we impose the following assumptions arising from a typical exposure process:
\begin{enumerate}
  % \begin{enumerate}[label = A.\arabic*]
\item The wave vector $\bm{k}$ and wave frequency $\omega$ of light sharply peak around $\bm{k}_{0}$ and $\omega_{0}$ (i.e., quasi-monochromatic) so that the investigated physical scale of the spacetime satisfies
\begin{align}\label{eq:con1}
\left\vert \Delta \bm{x} \right\vert \ll \left\vert \frac{1}{\Delta \bm{k}_{0}} \right\vert, \quad \left\vert \Delta t \right\vert \ll \left\vert \frac{1}{\Delta \omega_{0}} \right\vert
\end{align}
where $\Delta \bm{k}_{0} = \bm{k} - \bm{k}_{0}$ and $\Delta \omega_{0} = \omega - \omega_{0}$. The first condition in \eqref{eq:con1} reflects our interest in the spatially localized detection of light achieved by restricting the spatial integration domain in \eqref{eq:transMaster}. Therefore, unless otherwise specified, the spatial integration domain in \eqref{eq:transMaster} in the below is taken to be small enough to satisfy the condition in \eqref{eq:con1}.
\label{con:litho1}
        
\item The charged medium (i.e., the photoresist) is isotropic on the orthogonal plane of $\bm{k}_{0}$. For notational simplicity, we assume $\bm{k}_{0}$ to be along the $z$-direction so that $xy$-plane is the orthogonal plane of $\bm{k}_{0}$.
\label{con:litho2}
\end{enumerate}
    
The condition \ref{con:litho1} implies that the correlation of light in \eqref{eq:transMaster} is \cite{mandel1995}
\begin{align}
\left\langle \chi \right\vert A_{m}\left(x_{1}\right) A_{n}\left(x_{2}\right) \left\vert \chi \right\rangle \approx& \left\langle \chi \right\vert A^{\left( - \right)}_{m}\left(x_{0}\right) A^{\left( + \right)}_{n}\left(x_{0}\right)\left\vert \chi \right\rangle \times 2 \cos\left( k_{0}^{\mu}\left( x_{1} - x_{2} \right)_{\mu} \right),
\end{align}
where $k_{0}^{\mu} = \left( \omega_{0}, \bm{k}_{0} \right)$ for a spacetime point $x_{0} = \left(\bm{x}_{0}, t_{0}\right)$ in the integration domain of \eqref{eq:transMaster}.
In turn, \eqref{eq:transMaster} can be expressed as
\begin{align} \label{eq:byC1}
T\left(t, t_{0}\right) = \left\langle \chi \right\vert A_{\left( - \right)}^{m}\left(x_{0}\right) A_{\left( + \right)}^{n}\left(x_{0}\right)\left\vert \chi \right\rangle \kappa_{mn}
\end{align}
where 
\begin{align}
\kappa_{mn} = 2 \sum_{\substack{\psi, \bar{\psi} \in \mathcal{H}_{\psi}\\\psi\perp \bar{\psi}}} \int_{t_{0}}^{t} d^{4}x_{1} \int_{t_{0}}^{t} d^{4}x_{2} \left\langle \bar{\psi} \right \vert J_{m} \left(x_{1}\right) \left\vert \psi \right\rangle\left\langle \psi \right\vert J_{n} \left(x_{2}\right) \left\vert \bar{\psi} \right\rangle \cos\left( k_{0}^{\mu} \left( x_{1} - x_{2} \right)_{\mu} \right) .
\end{align}
Since $\kappa_{mn}$ is symmetric in its indices, the isotropy of the charged medium, together with the Coulomb gauge condition and condition \ref{con:litho1}, indicates that
\begin{align}
\kappa_{mn} = \delta_{mn} \eta, \quad \eta = \frac{1}{3} \text{Tr}\, \left( \kappa_{mn} \right).
\end{align}
Therefore, one finds
\begin{align}
T\left( t, t_{0} \right) = \eta \left\langle \chi \right\vert A^{m}_{(-)}\left( x_{0}\right) A^{(+)}_{m}\left( x_{0} \right) \left\vert \chi \right\rangle.
\end{align}
    
Finally, the optical equivalence theorem \cite{sudarshan1963, klauder1966} implies that $ A^{m}_{(-)}\left( x_{0} \right) A^{(+)}_{m}\left( x_{0} \right)$ 
is proportional to the quantum correspondence $I\left(x_{0}\right)$ of the classical light intensity. Absorbing the proportional factor
into $\eta$, we obtain the previously established photoelectric detection probability \cite{mandel1964}
\begin{align} \label{eq:detectionP}
p_{x_{0}} = \eta \left\langle I \left( x_{0}\right) \right\rangle.
\end{align}
where $\eta$ denotes the quantum efficiency.
    
Furthermore, by applying Poisson statistics to $p_{x_{0}}$, we can obtain the detection probability for $n$ photons in a spatial region $V$ and time interval $\Delta t$ as \cite{glauber1963}
\begin{align}\label{eq:poisson}
p_{\gamma, E = \left(V, \Delta t\right)}\left(n\right) = \left\langle \mathcal{T}\frac{{W}^{n}e^{-W}}{n!} \right\rangle
\end{align}
where $\mathcal{T}$ denotes the time ordering and 
\begin{align}
W = \int_{V} d^{3}x \int_{\Delta t} dt \eta \, \left(\bm{x}, t\right) I \left(\bm{x}, t\right).
\end{align}
corresponding to the total intensity absorbed by $\left(V, \Delta t\right)$.
The Poisson-like distribution $p_{\gamma, E}\left(n\right)$ in \eqref{eq:poisson} is often approximated as
\begin{align}
p_{\gamma, E}\left(n\right) \simeq \frac{\left\langle W\right\rangle^{n} e^{-\left\langle W \right\rangle}}{n!}
\end{align}
which reproduces the Poisson distribution of the number of photons in \cite{mack2010}.

\section{Analysis of probabilistic feature formation}
\label{sec:dep}

As outlined in the previous section, stochastic acid generation can be integrated with a chemical kinematic model to establish a mechanism for deprotection. In this section, we delve into a simple deprotection mechanism characterized by randomness arising exclusively from the photoelectric probability.

\subsection{Random deprotection density and probabilistic feature formation}

Given localized acid generation, the deprotection mechanism can be effectively depicted by a chemical kinetics model, as in \cite{houle2000}. In such a kinetics model, the density of the deprotected polymer at $\left( \bm{x}, t \right)$ arising from acid-catalyzed deprotection with respect to each localized acid generation at $\left( \bm{x}_{j}, t_{j} \right),\ j = 1, 2, \dots, n$ is given by $\rho_{D}\left(\bm{x}, t; \bm{x}_{j}, t_{j}\right)$, so that the total deprotection density $\rho_{D, \text{total}}$ is determined by \cite{gallatin2005}
\begin{align}
\rho_{D, \text{total}}\left( \bm{x}, t; \bm{x}_{j}, t_{j}, n \right) = \sum_{j = 1}^{n} \rho_{D}\left(\bm{x}, t; \bm{x}_{j}, t_{j} \right).
\end{align}

Here, $\rho_{D}$ denotes the diffusion of acid during the exposure process and subsequent post-exposure baking. The probability of photoelectric detection exclusively governs the stochastic nature of the diffusion. Practically, such diffusion is influenced by various base effects such as photo-decomposable quenchers (PDQs) and neutralization, which may introduce additional stochasticity to the deprotection density \cite{wang2010a}.

The quantum-mechanical description of acid generation in the previous section indicates that acid generation at a given spacetime occurs with probability $p_{\bm{x}, t}$ in \eqref{eq:detectionP}. Moreover, given a spacetime region $E$, one can easily find the probability distribution function for local generations of acid at $\left( \bm{x}_{j}, t_{j} \right),\ j = 1, 2, \dots, n$ as
\begin{align}
p_{\gamma}\left( \bm{x_{j}}, t_{j}, n \right) = p_{\gamma, E}\left( n \right) \prod_{k = 1}^{n} \hat{p}_{\gamma}\left( \bm{x}_{k}, t_{k} \right)
\end{align}
with $\hat{p}_{\gamma}\left( \bm{x}, t \right) = \frac{p_{\bm{x}, t}}{W}$ with the relative probability $p_{\bm{x}_{j}, t_{j}}$ of the acid generation at each $\left( \bm{x}_{j}, t_{j} \right)$. The total deprotection density $\rho_{D, \text{total}}\left(\bm{x}, t;\bm{x}_{j}, t_{j}, n\right)$ can then be regarded as a random field $\rho_{D, \text{total}}\left( \bm{x}, t; \bm{X}_{j}, T_{j}, N \right)$ over the spacetime derived from random variables
\begin{align}
\left( \bm{X}_{j}, T_{j}, N \right) \sim p_{\gamma}(\bm{x}_{j}, t_{j}, n)
\end{align}
where $\sim$ denotes the corresponding probability density function.

Using the relation between $p_{\bm{x}, t}$ and the aerial intensity $i_{\text{aerial}}\left(\bm{x}, t\right) = \left\langle I\left(\bm{x}, t\right)\right\rangle$ in \eqref{eq:detectionP}, we have the expectation of $\rho_{D, \text{total}}$ as
\begin{align} \label{eq:expectation}
&E\left[ \rho_{D, \text{total}}\left( \bm{x}, t \right) \right] 
= \left(\eta \rho_{D} \otimes i_{\text{aerial}}\right)\left(\bm{x}, t\right) \equiv \int d^{4}x \eta \rho_{D}\left( \bm{x}, t; \bm{x}', t' \right) i_{\text{aerial}}\left( \bm{x}', t'\right).
\end{align}
to which $\rho_{D, \text{total}}$ converges for a large absorption $W$ of open dose. The form \eqref{eq:expectation} implies that the expectation of $\rho_{D, \text{total}}$ corresponds to the convolution imaging of $i_{\text{aerial}}$ often taken as the nominal intensity $i_{\text{resist}}$ of the resist image.

In the patterning process, a pixel on a photoresist is formed if the density of the deprotected polymer exceeds a certain threshold $\tau$ at the development process \cite{gallatin2005, latypov2020}. Thus, the probability of generating a pixel at a given position $\bm{x}$ is given by
\begin{align} \label{eq:pixelP}
p_{\text{pixel}}\left( \bm{x} \right) = &P\left[ \rho_{D, \text{total}}\left( \bm{x}, t_{f} \right) \ge \tau \right]
\end{align}
where $P\left[\dots\right]$ denotes the probability of condition $\left[\dots\right]$ and $t_{f}$ is the final time for the exposure process. 

Similarly, a feature in the exposed region $E$ corresponds to a 3-dimensional object $P \in E$ such that for any $\bm{x} \in P$,
\begin{align}
    \rho_{D,\textrm{total}}(\bm{x}, t_{f}) \ge \tau.
% \left\{\bm{x} \in E \right. \left\vert \rho_{D, \text{total}}\left(\bm{x}, t_{f}\right) \ge \tau\right\}.
\end{align}
In turn, the probability of feature formation in a given region $P$ can be formally written as
\begin{align}\label{eq:featureP}
    p_{\textrm{feature}} = P\left[ \inf_{\bm{x} \in P} \rho_{D, \textrm{total}}(\bm{x}, t_{f}) \ge \tau \right].
\end{align}
Similarly, we may consider the probability of no feature formation in $P$:
\begin{align}\label{eq:missingP}
    p_{\textrm{missing}}(P) = 1 - P\left[ \sup_{\bm{x} \in P} \rho_{D,\textrm{total}}(\bm{x}, t_{f}) \ge \tau \right].
\end{align}
in which $P\left[ \sup_{\bm{x} \in P} \rho_{D,\textrm{total}}(\bm{x}, t_{f}) \ge \tau \right]$ is often called the excursion probability.

The formal probabilities in \eqref{eq:featureP} and \eqref{eq:missingP} enable us to define the probabilities of various printing failures. For instance, the probability of unexpected patterning (resp. missing) in a given region $P$ corresponds to $p_{\text{feature}}$ (resp. $p_{\text{missing}}$) \cite{latypov2020}. However, it should be noted that the computation of the excursion probability for a given random field is somewhat nontrivial, and its analytic formula has been known only for a few exceptional cases \cite{adler2009}.

\subsection{Asymptotic analysis of probabilistic feature formation}

We now discuss the asymptotic behavior of the probability density $\rho_{D, \text{total}}$ of a deprotected polymer for a large absorption $W$ in \eqref{eq:poisson}. A sufficiently large absorption can be achieved by increasing the exposure dose or the area of the exposed area of interest.
    
For a large $W$, only a large number $N$ of photoelectric interactions contributes significantly to the probability distribution of $\rho_{D, \textrm{total}}$. For such a large $N$, we can apply the central limit theorem so that
\begin{align}\label{eq:gaussianRF}
    \rho_{D, \text{total}}\left(\bm{x}, t_{f}\right) \sim \mathcal{N}\left( \frac{n}{W} i_{\text{resist}}\left(\bm{x}, t_{f}\right), n \sigma^{2}_{D}\left(\bm{x}, t_{f}\right)\right) p_{\gamma, E}(n)
\end{align}
where $\sigma^{2}_{D}$ denotes the variation in $\rho_{D}$ given by
\begin{align}
    \sigma^{2}_{D} = E\left[ \rho^{2}_{D}(\bm{x}, t_{f}) \right] - E\left[ \rho_{D}(\bm{x}, t_{f}) \right]^{2}
\end{align}
and $\mathcal{N}\left(\mu, \sigma^{2}\right)$ is a normal distribution with mean $\mu$ and variance $\sigma^{2}$.

By denoting the cumulative distribution function of the normal distribution in \eqref{eq:gaussianRF} as $\Phi$, we can obtain the probability of pixel formation at $\bm{x}$ as
\begin{align}
    p_{\textrm{pixel}}(\bm{x}) = \sum_{n = 0}^{\infty} \left( 1 - \Phi(\tau;n) \right) p_{\gamma,E}(n).
\end{align} 

As a simple example illustrating the stochastic analysis, we consider a family of static single-line patterns (i.e., the corresponding light intensity is constant in time) defined by aerial intensities in the form of
\begin{align}
i_{\text{aerial}}\left(x\right) = i_{0} \text{sinc}\,\left(2 \pi x\right)^{2}, \quad \text{sinc}\,\left(x\right) = \frac{\sin\left(x\right)}{\texttt{}x},
\end{align}
which are classified by the relative intensity $i_{0}$ at $x = 0$. 

\begin{figure}
  \centering
  \scalebox{0.5}{\input{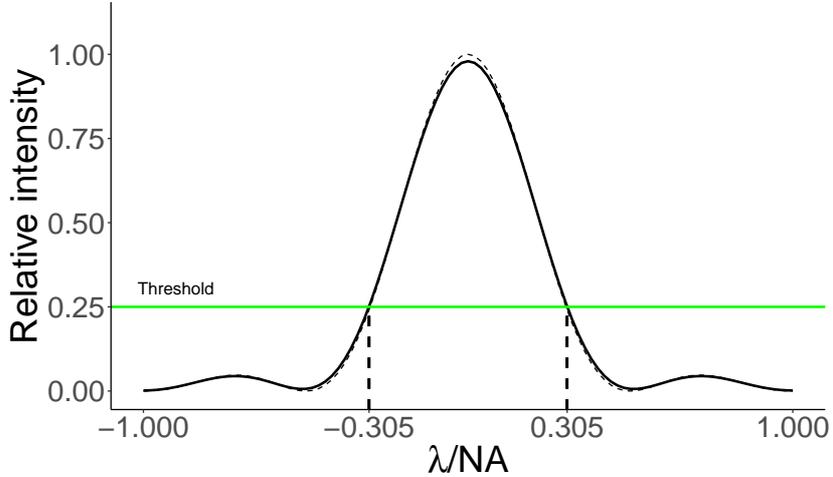}}
  \caption{The simplest family of single-line patterns defined by aerial intensity. The dashed and solid lines indicate the aerial and resist image intensities, respectively. The interval determined by the threshold (green line) corresponds to the nominally patterned region.} 
  \label{fig:slIntensity}
\end{figure}

For simplicity, we set the quantum efficiency $\eta = 1$ and employ a normalized Gaussian kernel for the deprotection kernel, that is,
\begin{align}
\rho_{D}\left(x\right) = \mathcal{N}\left(x, h^{2}\right)
\end{align}
in which $h = 1/50$ satisfies our purpose here. Subsequently, given the threshold $\tau$, the critical dimension (CD) of a single line is determined by the interval $\left\{x \in \mathbb{R} \right.\left\vert \rho_{D, \text{total}}\left(x\right) \ge \tau\right\}$ (see Figure \ref{fig:slIntensity}).

\begin{figure}
  \centering
  \scalebox{0.5}{\input{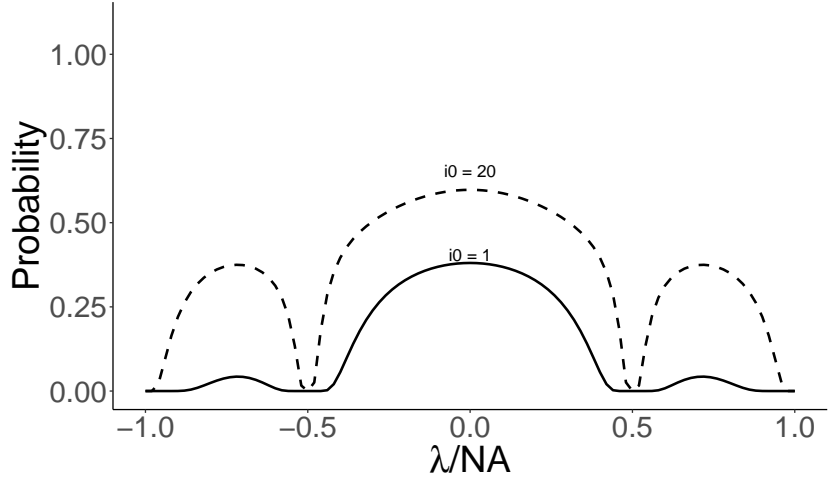}}
  \caption{Probabilities for a single-pixel formation of a single line under two relative intensities.} 
  \label{fig:pixelProb}
  \end{figure}

For a large $i_{0}$, we may approximate $\rho_{D,\textrm{tottal}}$ as in \eqref{eq:gaussianRF}.
Figure \ref{fig:pixelProb} illustrates the probability of pixel generation at each point on $x$ with two different relative intensities: $i_{0} =80$ and $i_{0} = 120$.

  \begin{figure*}[h]
    \centerline{\scalebox{0.4}{\input{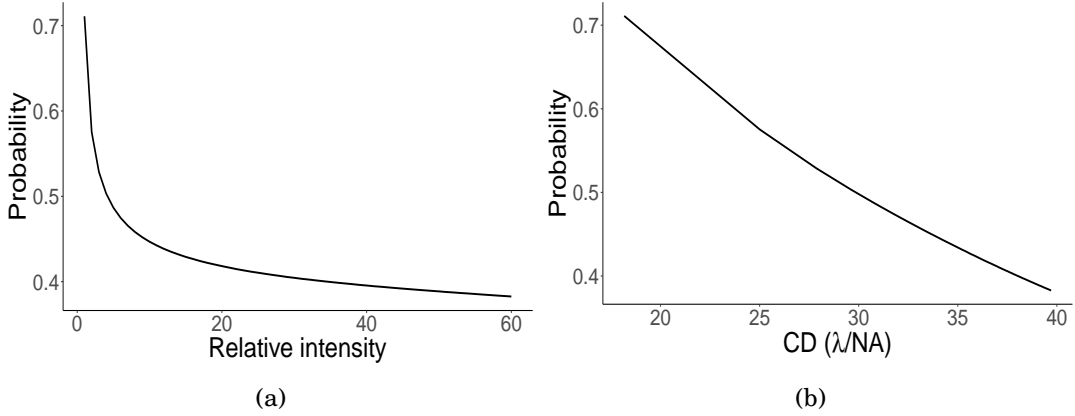}}}
    
    \caption{probability of no pattering in the region reserved by the nominal resist intensity with respect to the relative dose (left) or nominal CD (right).} 
    \label{fig:missing}
  \end{figure*}

Finally, we consider the pattern-missing probability for a nominally patterned region. As noted above, the exact computation of excursion probability is highly demanding. However, the lower bound of an excursion probability can be computed if a unique peak in $P$ exists in the second moment of the Gaussian random field. Explicitly, 
\begin{align}
    P\left[ \sup_{\bm{x} \in P} \rho_{D,\textrm{total}}(\bm{x}, t_{f}) \ge \tau \right] \ge \Phi\left( \frac{\tau}{\left( \sup_{x \in P} E\left[ \rho_{D,\textrm{total}}^{2}(x) \right] \right)^{1/2}} \right).
\end{align}
The lower bound of the excursion probability on $P$ then provides the upper bound of the missing probability as
\begin{align}
    p_{\textrm{missing}}(P) \le 1 - \Phi\left( \frac{\tau}{\left( \sup_{x \in P} E\left[ \rho_{D,\textrm{total}}^{2}(x) \right] \right)^{1/2}} \right).
\end{align}

Figure \ref{fig:missing} shows the relation between the probability of no patterning in the nominally patterned area and the relative intensity (left) or nominal CD (right). As shown in both figures, our simple example qualitatively captures the characteristics of the stochastic printing failures in \cite{bisschop2017}.

Another consequence of a large $W$ condition is the probabilistic deviation of the boundary of a given feature from its nominal boundary. For this, let $\bm{x}_{0}$ be the nominal boundary of the feature, so that for a given threshold $\tau$, $E\left[\rho_{D, \text{total}}\left(\bm{x}_{0}\right)\right] = \tau$. 

For a large $W$ condition, we may consider the probabilistic deviation of the boundary of a given feature from its nominal boundary. For this, let $\bm{x}_{0}$ be the nominal boundary of the feature, so that for a given threshold $\tau$, $E\left[\rho_{D, \text{total}}\left(\bm{x}_{0}\right)\right] = \tau$. 
    
Then for a small deviation $\delta \bm{x}$ we have the condition
\begin{align}
\tau =& \rho_{D, \text{total}}\left(\bm{x}_{0} + \delta \bm{x}, t ; \bm{X}_{j}, T_{j}\right)\\
\simeq& 
\rho_{D, \text{total}}\left(\bm{x}_{0} , t ; \bm{X}_{j}, T_{j}\right) + 
\delta x^{m} \partial_{m} \rho_{D, \text{total}}\left(\bm{x}_{0}, t ; \bm{X}_{j}, T_{j}\right) \notag
\end{align}
to $\bm{x}_{0} + \delta \bm{x}$ be a deviated boundary. Therefore, by defining the line edge roughness (LER) as
\begin{align}
\text{LER} = E\left[\delta x^{2}\right]
\end{align}
one finds
\begin{align}
\text{LER} \simeq \frac{W \sigma_{D}^{2}}{\partial^{m} i_{\text{resist}}\left(\bm{x}_{0}\right)\partial_{m} i_{\text{resist}}\left(\bm{x}_{0}\right)}
\end{align}
which reproduces the result of \cite{gallatin2005}.

\section{Conclusion and outlook}
\label{sec:con}

In this work, we have applied QED to obtain a fully quantum-mechanical description of photochemical acid generation identified with the photoelectric detection of light. We have seen that the well-known proportionality between the probability of acid generation and the intensity of incident light can be reproduced under certain assumptions in the QED framework.
	
Although we have focused merely on the occurrence of acid generation, a more detailed analysis is available in the QED framework by emphasizing the dynamics of charged matter and the final state of the exposure system. For instance, one may split the final state of the photons by the momentum of a state having an acute angle with respect to the incident wave vector to investigate the rate of flare generation during exposure. Also, by specifying the dynamics of charged matter to be that of electrons through the spin 1/2-field, one can determine the rate of generation of secondary electrons having sufficient energy to further trigger the deprotection mechanism. In this sense, the QED formulation provides an integrated framework for investigating secondary effects, which will be discussed in future studies.
	
Using probabilistic acid generation, we have constructed a probabilistic description of deprotection density in the photoresist, as reported by \cite{gallatin2005,latypov2020}. The probabilistic nature of the deprotection density reveals the relationship between the expectation of deprotection and the resist image intensity in convolution imaging. 

In addition, the explicit form of the deprotection density allows us to perform an asymptotic analysis under a large absorption condition, which refines the work of \cite{gallatin2005,latypov2020}. In particular, under asymptotic conditions, we have found that deprotection density can be depicted by a Gaussian random field, enabling the simulation of randomness in the resist image using a Monte Carlo type simulation.
	
In a typical lithography process, a large absorption condition can be achieved by enlarging the exposed area of interest, and an asymptotic analysis can thus be used to construct a simulation model for quantum fluctuations in the patterning process. In particular, by approximating the deprotection blur $\kappa = \left(\int_{E} d^{4}x \eta\right) \times \left(\int_{E} d^{4}y \rho^{2}_{D}\left(y\right)\right)$ as a constant, we can express the asymptotic formula as follows:
\begin{align}
\rho_{D, \text{total}}\left(\bm{x}, t_{f}\right) \sim \mathcal{N}\left(i_{\text{resist}}\left(\bm{x}, t_{f}\right), \kappa i_{\text{aerial}}\left(\bm{x}, t_{f}\right)\right)
\end{align}
so that $\kappa$ is fitted by measurement under the given resist image intensity and aerial intensity computed by a conventional simulation model. This simulation model is conceptually similar to that in \cite{latypov2020}, which has been widely applied to construct EUV stochastic models \cite{latypov2022,wei2022,pan2023,tsai2024}.
	
Finally, the parameter $\kappa$ can be considered as the leading characteristic of the photoresist, which governs the sensitivity of the resist under quantum fluctuations. Thus, $\kappa$ can be considered a physically refined parameter of the $k_{4}$ factor in \cite{santaclara2020}.

%\backmatter

\section*{Acknowledgments}
SL expresses gratitude to Dawoon Choi and Yunkyoung Song for their enlightening discussions during the initial phase of this work. This work was partly supported by the Institute for Information \& Communications Technology Promotion (IITP) grant (No. 2019-0-00003, Research and Development of Core Technologies for Programming, Running, Implementing, and Validating of Fault-Tolerant Quantum Computing System) and National Research Foundation of Korea (NRF) grants (Nos. 2024M3K5A1004355, RS-2024-00432214, RS-2023-00281456, and RS-2023-00283771). It's worth noting that this work was partially completed at Samsung Electronics, one of the author's former affiliations.

\bibliographystyle{jhep} 
\bibliography{stochastic_optics_arxiv.bib}%

\providecommand{\href}[2]{#2}\begingroup\raggedright\begin{thebibliography}{10}

\bibitem{mack2010}
C.A.~Mack, ``Line-edge roughness and the ultimate limits of lithography,''
  \href{https://doi.org/10.1117/12.848236}{\emph{Advances in {{Resist
  Materials}} and {{Processing Technology XXVII}}}, vol.~7639, 901}, {SPIE},
  (2010).

\bibitem{schwinger1948}
J.~Schwinger, ``Quantum {{Electrodynamics}}. {{I}}. {{A Covariant
  Formulation}},'' \href{https://doi.org/10.1103/PhysRev.74.1439}{\emph{Phys.
  Rev.} {\bfseries 74} (1948) 1439}.

\bibitem{mandel1964}
L.~Mandel, E.C.G.~Sudarshan and E.~Wolf, ``Theory of photoelectric detection of
  light fluctuations,''
  \href{https://doi.org/10.1088/0370-1328/84/3/313}{\emph{Proc. Phys. Soc.}
  {\bfseries 84} (1964) 435}.

\bibitem{gallatin2005}
G.M.~Gallatin, ``Resist blur and line edge roughness,'' {\emph{Optical
  {{Microlithography XVIII}}}, vol.~5754, 38}, {SPIE}, (2005).

\bibitem{latypov2020}
A.~Latypov, G.~Khaira, G.~Fenger, J.~Sturtevant, C.-I.~Wei and P.D.~Bisschop,
  ``Probability prediction of {{EUV}} process failure due to resist-exposure
  stochastic: Applications of {{Gaussian}} random fields excursions and
  {{Rice}}'s formula,'' \href{https://doi.org/10.1117/12.2551965}{\emph{Extreme
  {{Ultraviolet}} ({{EUV}}) {{Lithography XI}}}, vol.~11323, 140}, {SPIE},
  (2020).

\bibitem{weinberg1995}
S.~Weinberg, \emph{The {{Quantum Theory}} of {{Fields}}: {{Volume}} 1,
  {{Foundations}}}, {Cambridge University Press} (1995).

\bibitem{messiah2014}
A.~Messiah, \emph{Quantum {{Mechanics}}}, {Dover Publications} (2014).

\bibitem{mandel1995}
L.~Mandel and E.~Wolf, \emph{Optical {{Coherence}} and {{Quantum Optics}}},
  {Cambridge University Press} (1995).

\bibitem{sudarshan1963}
E.C.G.~Sudarshan, ``Equivalence of {{Semiclassical}} and {{Quantum Mechanical
  Descriptions}} of {{Statistical Light Beams}},''
  \href{https://doi.org/10.1103/PhysRevLett.10.277}{\emph{Phys. Rev. Lett.}
  {\bfseries 10} (1963) 277}.

\bibitem{klauder1966}
J.R.~Klauder, ``Improved {{Version}} of {{Optical Equivalence Theorem}},''
  \href{https://doi.org/10.1103/PhysRevLett.16.534}{\emph{Phys. Rev. Lett.}
  {\bfseries 16} (1966) 534}.

\bibitem{glauber1963}
R.J.~Glauber, ``The {{Quantum Theory}} of {{Optical Coherence}},''
  \href{https://doi.org/10.1103/PhysRev.130.2529}{\emph{Phys. Rev.} {\bfseries
  130} (1963) 2529}.

\bibitem{houle2000}
F.A.~Houle, W.D.~Hinsberg, M.~Morrison, M.I.~Sanchez, G.~Wallraff, C.~Larson
  et~al., ``Determination of coupled acid catalysis-diffusion processes in a
  positive-tone chemically amplified photoresist,''
  \href{https://doi.org/10.1116/1.1303753}{\emph{J. Vac. Sci. Technol. B}
  {\bfseries 18} (2000) 1874}.

\bibitem{wang2010a}
C.W.~Wang, C.Y.~Chang and Y.~Ku, ``{Photobase generator and photo decomposable
  quencher for high-resolution photoresist applications},''
  \href{https://doi.org/10.1117/12.848623}{\emph{Advances in Resist Materials
  and Processing Technology XXVII}, vol.~7639, 76390W}, International Society
  for Optics and Photonics, SPIE, (2010).

\bibitem{adler2009}
R.J.~Adler and J.E.~Taylor, \emph{Random {{Fields}} and {{Geometry}}}, Springer
  (2009).

\bibitem{bisschop2017}
P.D.~Bisschop, ``Stochastic effects in {{EUV}} lithography: Random, local
  {{CD}} variability, and printing failures,''
  \href{https://doi.org/10.1117/1.JMM.16.4.041013}{\emph{JM3.1} {\bfseries 16}
  (2017) 041013}.

\bibitem{latypov2022}
A.~Latypov, C.-I.~Wei, P.D.~Bisschop, G.~Khaira and G.~Fenger, ``Calibration of
  {{Gaussian}} random field stochastic {{EUV}} models,''
  \href{https://doi.org/10.1117/12.2614142}{\emph{Optical and {{EUV
  Nanolithography XXXV}}}, vol.~12051, 28}, SPIE, (2022).

\bibitem{wei2022}
C.-I.~Wei, A.~Latypov, P.D.~Bisschop, G.~Khaira and G.~Fenger, ``Calibration
  and application of {{Gaussian}} random field models for exposure and resist
  stochastic in {{EUV}} lithography,''
  \href{https://doi.org/10.35848/1347-4065/ac54f5}{\emph{Jpn. J. Appl. Phys.}
  {\bfseries 61} (2022) SD0806}.

\bibitem{pan2023}
Z.~Pan, A.~Latypov, C.-I.~Wei, P.D.~Bisschop, G.~Fenger and J.~Sturtevant,
  ``Importance sampling in {{Gaussian}} random field {{EUV}} stochastic model
  for quantification of stochastic variability of {{EUV}} vias,''
  \href{https://doi.org/10.1117/12.2658651}{\emph{Optical and {{EUV
  Nanolithography XXXVI}}}, vol.~12494, 425}, SPIE, (2023).

\bibitem{tsai2024}
Y.-P.~Tsai, C.-M.~Chang, Y.-H.~Chang, A.~Oak, D.~Trivkovic and R.-H.~Kim,
  ``{Study of EUV stochastic defect on wafer yield},''
  \href{https://doi.org/10.1117/12.3010858}{\emph{DTCO and Computational
  Patterning III}, vol.~12954, 1295404}, International Society for Optics and
  Photonics, SPIE, (2024).

\bibitem{santaclara2020}
J.G.~Santaclara, B.~Geh, A.~Yen, T.A.~Brunner, D.D.~Simone, J.~Severi et~al.,
  ``One metric to rule them all: New k4 definition for photoresist
  characterization,'' \href{https://doi.org/10.1117/12.2554493}{\emph{Extreme
  {{Ultraviolet}} ({{EUV}}) {{Lithography XI}}}, vol.~11323, 321}, {SPIE},
  (2020).

\end{thebibliography}\endgroup

\end{document}